Perspective

# Call for a Framework for Reporting Evidence for Life Beyond Earth

James Green[1], Tori Hoehler[2], Marc Neveu[3], Shawn Domagal-Goldman[4], Daniella Scalice[5], Mary Voytek[6]

Submitted to *Nature* February 5, 2021

[1]Office of the Chief Scientist, NASA Headquarters, 300 Hidden Figures Way SW, Washington DC 20546 USA, Corresponding Authors email: james.green@nasa.gov

[2]Exobiology Branch, NASA Ames Research Center, P.O. Box 1, Moffett Field, CA 94035 USA

[3]Planetary Environments Laboratory, NASA Goddard Space Flight Center, Greenbelt, MD 20771 USA
Department of Astronomy, University of Maryland, College Park, MD 20742 USA

[4]Planetary Systems Lab, NASA Goddard Space Flight Center, 8800 Greenbelt Road, Greenbelt, MD 20771 USA

[5]NASA Astrobiology Program, NASA Ames Research Center/Paragon TEC, Moffett Field, CA 94035 USA

[6]Planetary Science Division, NASA Headquarters, 300 Hidden Figures Way SW, Washington DC 20546 USA

## Preface

Ours could realistically be the generation to discover evidence of life beyond Earth. With this privileged potential comes responsibility. The magnitude of the question, "are we alone?", and the public interest therein, opens the possibility that results may be taken to imply more than the observations support, or than the observers intend. As life detection objectives become increasingly prominent in space sciences, it is essential to open a community dialog about how to convey information in a subject matter that is diverse, complicated, and has high potential to be sensationalized. Establishing best practices for communicating about life detection can serve to set reasonable expectations on the early stages of a hugely challenging endeavor, attach value to incremental steps along the path, and build public trust by making clear that 'false starts' and 'dead ends' are an expected and potentially productive part of the scientific process. Here, we endeavor to motivate and seed the discussion with basic considerations and offer an example of how such considerations might be incorporated and applied in a proof-of-concept-level framework. Everything mentioned herein, including the name of the confidence scale, is intended not as a prescription, but simply as the beginning of an important dialogue.

______________________________________________________________________

The question "are we alone in the universe?" has been a source of wonder for humanity for millennia or more (1). Advances in planetary sciences, astronomy, biology, and other fields now leave us poised, as never before, to address this question with scientific rigor. Numerous challenges confront this endeavor, including challenges of perception and communication. Evidence of life may be subtle or unfamiliar, and reveal itself only in stages, as one observing campaign informs the next. However, the search for such evidence is often framed as an all-or-nothing proposition: either a mission returns definitive evidence of life or it has fallen short of its objective. The binary nature of this framing poses significant risk to the overall endeavor by levying unrealistically high expectations on its initial stages. This is true in communicating both the potential science yield of proposed observations and the ultimate results of those observations. It devalues science that yields progress along the path but falls short of definitive life detection, and risks eroding public confidence if reports of life detection are later shown to be ambiguous or inaccurate. Indeed, history includes many claims of life detection that later proved incorrect or ambiguous (2) when considered in exclusively binary terms. If, instead, we recast the search for life as a progressive endeavor, we convey the value of observations that are contextual or suggestive but not definitive and emphasize that false starts and dead ends are an expected part of a healthy scientific process.

Realizing this potential requires a community-level dialog among scientists, technologists, and the media to agree on objective standards of evidence for life and best practices for communicating that evidence. Doing so before life detection results are reported, rather than in response to a specific finding, will enable a more dispassionate, objective, and broad-reaching treatment of the subject and ensure that its packaging supports clear understanding by the public at large. The purpose of this paper is to call for such dialog and propose a draft framework and set of considerations to seed the discussion.

## Considerations for a Progressive Scale

Communication within and from the scientific community should "Inform, not persuade; Offer balance, not false balance; Disclose uncertainties; State evidence quality; and Inoculate against misinformation" (3). While relevant across all scientific disciplines, this guidance is particularly meaningful in communicating about a subject matter that is complex, diverse, and has high potential to be misunderstood, oversold, or sensationalized. We focus in particular on evidence quality and uncertainty as critical considerations.

The National Academies report, "An Astrobiology Strategy for the Search for Life in the Universe" (4) proposed that the utility of various life detection measurements be evaluated primarily with respect to their potential to yield false negative and/or false positive interpretations[1]. This framing underscores the need to clearly convey the *type* of information that a given measurement can provide, which we suggest could be described as a balance of stringent, inclusive, and contextual components. *Stringent* measurements emphasize the avoidance of false positive results by seeking features that are strongly indicative of life, but possibly at the expense of overlooking some indicators of life. *Inclusive* measurements emphasize the avoidance of false negative results by seeking features that are potentially indicative of life, but possibly at the expense of providing definitive evidence. *Contextual* measurements are not inherently indicative of life but are important for interpreting the results of measurements that may be.

Where a given measurement falls along this spectrum should not be seen to indicate its 'value', but rather its utility in application to a particular target at a particular time, given the existing state of knowledge and in light of constraints on the mission. To be objective, the classification of measurements according to these or other metrics should ideally occur via discussion within the community *before* they are deployed to seek evidence of life. The 'Ladder of Life Detection' (5) took the initial steps in this process. Clear identification of information *type* is important in formulating life detection objectives, expressing the potential inherent in specific measurements or mission architectures, and communicating results to the scientific community, the media, and the general public. For each of these audiences, it is also essential to express the degree of confidence or uncertainty with which results are reported. While it is common practice in science to convey the uncertainty associated with specific measurements, we propose that this should also be clearly articulated for the *overall interpretation* and conveyed at all levels of communication.

The diverse combinations of measurements that could be made to seek evidence of life implies that a comparably diverse set of factors will govern the potential for false positive and false negative results, and also the level of certainty with which these results can be interpreted. However, an effort to represent this diversity in a standardized way – for example, as a uniform set of factors that are addressed in any communication relating to life detection – can enable clearer understanding by a broad audience. Doing so ensures that communication of potential or results clearly addresses not only what can be determined by a given set of measurements, but also what cannot be. In the context of life detection, the latter is just as important as the former in accurately conveying the significance of a series of planned or executed measurements.

As a starting point for discussion, we propose that a standard set of information conveyed should minimally address: Individual measurement quality, including signal-to-noise ratio, replicability, potential for ambiguity (to what extent the methodology is specific to the targeted observable or subject to interferences), and discussion of how direct or derived the reported results are relative to the measured parameter; the potential contribution from Terrestrial contamination, which directly affects the potential for false positive results; the potential contribution from abiotic sources, which directly affects the potential for false positive results; the congruence (goodness of fit) of a biological explanation given the environmental context; and whether the overall interpretation is supported by independent lines of evidence or measurement ensembles. "Independent" may refer to differing analytes (distinct potential biosignatures), analytical approaches, or mission platforms. "Ensembles" refers to measurements made on multiple target bodies – for example, multiple exoplanets – in order to build evidence by identifying statistical trends over a large population of objects (6).

A particular challenge in life detection is that interpretation may depend to varying degrees on environmental context that may, in turn, be established to a greater or lesser extent. Discussion within each of the above categories should clearly identify any such dependency and its propagation into the overall confidence with which results are reported. Similarly, reporting should clearly indicate whether the interpretation of evidence is tied to a specific definition of life or conception of its unique attributes, and development of a standardized framework for evidence communication should account for the possibility that such definitions and conceptions may vary (7).

Dialog within the scientific community may augment or replace these categories. Whatever consensus is reached should represent a template for communication that acknowledges the importance of objectively assessing what can *and* cannot be said, and with what confidence, in each standard category.

While the range of factors discussed above is diverse and, in a sense, multi-dimensional, it is nevertheless desirable to seek a means of capturing this information in a concise metric that clearly and understandably communicates the significance of a new result. We have also argued that it is essential to convey that life detection results are a matter of degree, not a purely yes-or-no proposition. To reconcile these two perspectives, we propose that the scientific community develop a progressive one-dimensional scale that clearly communicates where a given objective or result falls within the spectrum of the overall life detection endeavor.

---

[1] False negative results are conveyed when a set of measurements fails to indicate that life is or was, in fact, present. False positive results are conveyed when a set of measurements are interpreted as evidence of life when life is or was *not*, in fact, present.

## An Example Progressive Scale

A number of progressive scales have been developed. For example, the Torino Scale (8) was designed to convey the potential impact hazard from a close approaching Near Earth Object (NEO), another topic with potential to be sensationalized. This approach distilled a complicated probability calculation along with individual NEO characteristics and estimates of destructiveness into a four-color scale (red, orange, yellow, green) that was designed to communicate a level of concern. Notably, the scale evolved since its initial formulation in order to more directly convey to the public the implications of its levels (9). Since NEO Earth impacts are extremely rare, there has been no opportunity to effectively use the scale as designed.

NASA utilizes a similar approach, the Technological Readiness Level (TRL) scale, to characterize the maturation of instruments from concept to implementation in flight (10). Instruments used in spaceflight vary widely in their principles of operation, development pathways, approaches to benchmarking, and even the language used to describe them. While the full set of information needed to describe the process of maturation is complex and varied, it has nevertheless proven invaluable to track the development of diverse instruments relative to a standard set of benchmarks, along the one-dimensional progressive TRL scale (Table 1). Progression along the scale concisely conveys the state of maturity and increases confidence that a given technology can be successfully implemented in flight, in a fashion that is applicable across a diversity of instrument types.

| TRL | System Characteristics |
|---|---|
| 9 | Actual system flight proven through successful mission operations |
| 8 | Actual system flight qualified through test and demonstration |
| 7 | System prototype demonstration in a space environment |

| 6 | System/subsystem prototype demonstrated in a relevant environment |
| 5 | Assembly/component brass-board validation in a relevant environment |
| 4 | Assembly/component brass-board validation in a laboratory environment |
| 3 | Analytical and/or experimental performance/function proof of concept |
| 2 | Technology concept and/or application formulated |
| 1 | Basic principles observed and reported |

Table 1. Progressive scale for NASA's Technology Readiness Levels and their meaning (adapted from (10)).

By analogy to the TRL and Torino scales, we propose that a one-dimensional progressive scale for characterizing the status of a life detection investigation will serve to convey information at a balanced level: more than simply 'yes' or 'no', but with a reduced complexity that enables a generalist audience to quickly understand the significance of a new result. Such a scale has potential for application at many levels: in conveying the current understanding of a given target of exploration as it bears on life detection; in formulating new objectives; in accurately conveying the potential represented in a proposed instrument capability, suite of measurements, or mission implementation; and in communicating results to the scientific community, media, and public. As such, the scale should be developed through dialog among the full set of stakeholders in this endeavor, including a broad cross section of scientists, instrumentalists, engineers, and communication experts. Here, we endeavor to both motivate this discussion and seed it with a non-prescriptive example of how the considerations discussed above could be condensed into a one-dimensional "Confidence of Life Detection" ("CoLD") scale.

In the example CoLD scale (Fig. 1), the progression in confidence that a set of observations stands as evidence of life is marked by seven benchmarks. The lowest levels of this scale focus on initial identification of potential biosignatures – for example, chemistry, physical structures, or activity consistent with biological origin. The intermediate levels involve establishing the physical and chemical context of the environment required to assess habitability, the congruence of a biological explanation, and the potential for abiotic or contaminant sources of the potential biosignature. Higher levels of the scale involve corroboration of the initial result by independent lines of evidence and dismissal of alternative hypotheses that are developed by the community specifically in response to the initial result. Exploration and implementation of statistical methods may be necessary to determine the degree of confidence with which each set of measurement indicators can be achieved for each level of the CoLD scale. This is particularly relevant for Level 4, as ruling out all possible non-biological signals may be a significant challenge, especially for bodies/environments with remote sensing data only. Achieving Level 7 results would involve a follow-up investigation, perhaps including a dedicated mission, planned after achieving Level 6 results.

Three important concepts are inherent in such a scale: 1) results conveyed objectively at any level of this scale have value in the overall life detection endeavor; 2) it is a natural and expected outcome of the scientific process that individual results may not progress beyond a certain level, and may move to lower levels as those results are critically evaluated (for example: sources of contamination subsequently determined); and 3) achieving the highest levels of confidence requires the active participation of the broader scientific community.

## Example Applications of the CoLD Scale

To exemplify how a progressive scale can aid in communicating both results and the potential inherent in observations yet to be made, we briefly consider applications to the report of evidence for life in the Martian meteorite ALH84001, the scientific potential of an active mission (the Mars Perseverance rover), and the scientific potential of mission concepts to seek evidence of life on exoplanets.

The report of evidence for life in the 4.5 billion-year old Martian meteorite ALH84001 (11) illustrates challenges inherent in communicating results in a binary 'yes or no' fashion and the benefits of a progressive scale. ALH84001 contains carbonate globules and associated magnetite particles argued to have properties consistent with microbially-induced formation, and carbonate-associated polycyclic aromatic hydrocarbons (PAHs) – possible biomolecule remnants. It also contains co-occurring magnetite and iron sulfides, seen as compatible with microbial oxidation-reduction processes. Analysis indicated each of these features to be indigenous to ALH84001 and it was argued that they formed in liquid water at 0°–80°C (12) at a time when the Martian surface was likely habitable. McKay et al. (11) acknowledged that each potentially biogenic feature could also have an abiotic explanation but concluded that their observations collectively evidenced life on early Mars. The publication and public announcement of these results led to efforts by the broader community to develop a more extensive case for abiotic origin of the features as a collective set (13). As a result, the present consensus is that the features identified by McKay et al. (11) in ALH84001 are probably not indicative of life.

The binary nature of the researchers' conclusions (life) and the community response (not life) cast the report in the light of a false and refuted claim rather than as a useful part of a larger progression. However, the report by McKay et al. (11) and the ensuing community response significantly advanced the search for life beyond Earth via an increasingly detailed and rigorous understanding of what constitutes evidence for life. In the framework proposed here, when measured against a progressive scale, the observations could be argued to have identified features known to result from biological processes (CoLD Level 1), ruled out contamination (Level 2), and established the congruence of a biological explanation (Level 3), but clearly not to have met the higher benchmarks of the scale. As such, it would stand as useful progress that motivates follow up work but not as definitive evidence of life.

A confidence scale can also be applied to assess the potential of a suite of observations yet to be made, which is valuable both in weighing alternative strategies and in accurately setting expectations. For example, the Mars 2020 Perseverance Rover will seek potential signs of ancient life via *in situ* observations at Jezero Crater and cache samples as a first step toward return of Martian surface materials to Earth (14, 15). Jezero Crater hosts carbonates in a former lake and its tributaries. On Earth, lacustrine carbonates tend to preserve morphological, organic, and isotopic biosignatures (e.g., 16, 17, 18, 19), and can be biologically mediated (e.g., 20).

The *in situ* investigation of these features could gather evidence that reaches CoLD Level 3: the rover's instruments could detect potentially biogenic features (Level 1) of a nature that would not definitively rule out abiotic explanations (< Level 4); controls can be utilized to assess the potential for terrestrial contamination (Level 2); and the onboard instrument suite can make measurements suitable to establish congruent/habitable environmental context (Level 3). Return of samples to Earth could yield evidence that reaches Level 6 or higher: extensive analysis using a range of techniques could establish the detailed context needed to develop and test abiotic hypotheses for potential biosignatures (Level 4); observations can be validated by independent investigators who develop and follow up on new hypotheses (Level 5); and the diversity of current and future analytical techniques that could be brought to bear has potential to yield multiple lines of evidence independent from what Perseverance can identify (Level 6), such as specific organic compounds (19) and their isotopic composition (21). Achieving Level 7 may require investigations elsewhere on Mars where the above results would predict specific signs of life.

A progressive scale could also be applied to mission conceptualization, for example in the flagship telescope mission concepts studied as inputs to the Decadal Survey on Astronomy and Astrophysics 2020 (22). Three of these concepts - HabEx, LUVOIR, and Origins - incorporated a search for evidence of life on exoplanets into their goals. While the example framework discussed here did not exist during the study of these concepts, a similar set of considerations is nevertheless evident in their formulation. Each of these missions would search for global biospheres through the detection of gases that can be biologically produced (CoLD Level 1). Using observing geometries that would limit reflection of light from Earth off spacecraft elements, they could nearly eliminate signal contamination from Earth (CoLD Level 2). All would constrain the environments of their target planets - Origins by studying the temperature profile of the planet's atmosphere and HabEx and LUVOIR by searching for evidence of liquid water oceans (CoLD Level 3). These studies incorporated research (see reference 23 for a review) on pathways for abiotic production of $O_2$ (a potential biosignature gas and key target) - and designed observations that would discriminate between abiotic and biological sources (CoLD Level 4), for example by detecting methane ($CH_4$) in addition to molecular oxygen ($O_2$) or ozone ($O_3$). The LUVOIR study considered ways to search for spectral features of surface pigments that would provide a second, independent signal of life; as could HabEx in ideal cases (CoLD Level 5). The LUVOIR study also prioritized the ability to search for signs of life on dozens of rocky, Earth-sized exoplanets. This sample size would enable two additional applications of the CoLD scale: first, it would allow for meaningful tests that would search for signs of life on planets with/without markers of habitability; and it would allow for a statistical assessment of the number of planets (e.g., 24) that meet any level of the CoLD scale. Achieving CoLD Level 7, which is related to follow-up observations and tests, would likely necessitate second-generation instruments or missions, or long-term observations.

## A Call to Action

The CoLD scale and example applications are offered to demonstrate intent, proof of concept, and utility, and are not a prescription. Discourse within the broader community should modify or supplant the scale to ensure its applicability across the diversity of targets and methods that may become part of the search for evidence of life beyond Earth. Indeed, much remains to be established beyond the basic framework offered here. Is a one-dimensional progressive scale the right approach or are additional 'axes' needed? What specific 'standards' should be included and in what progression? How does (un)certainty in separate lines of evidence propagate into overall confidence, and how should ensemble/probabilistic approaches be encompassed? How do we ensure sufficient flexibility to accommodate the very different sorts of observations that will be made to seek evidence of life in the diverse environments of our solar system and the myriad worlds beyond? Whatever the outcome of the dialog, what matters is that it occurs: that the community works together, now, to develop objective standards for evidence communication. In doing so, we can only become more effective at communicating the results of our work, and the wonder associated with it, in a way that captures our own uncertainties and passion for those whose work will follow.

**Acknowledgements:** We acknowledge the excellent discussion and suggested changes made by the editor and the reviewers which have improved the paper.

**Author contributions:** All authors contributed to writing the manuscript after many rigorous discussions.

**Competing interests:** The authors declare they have no competing interests.

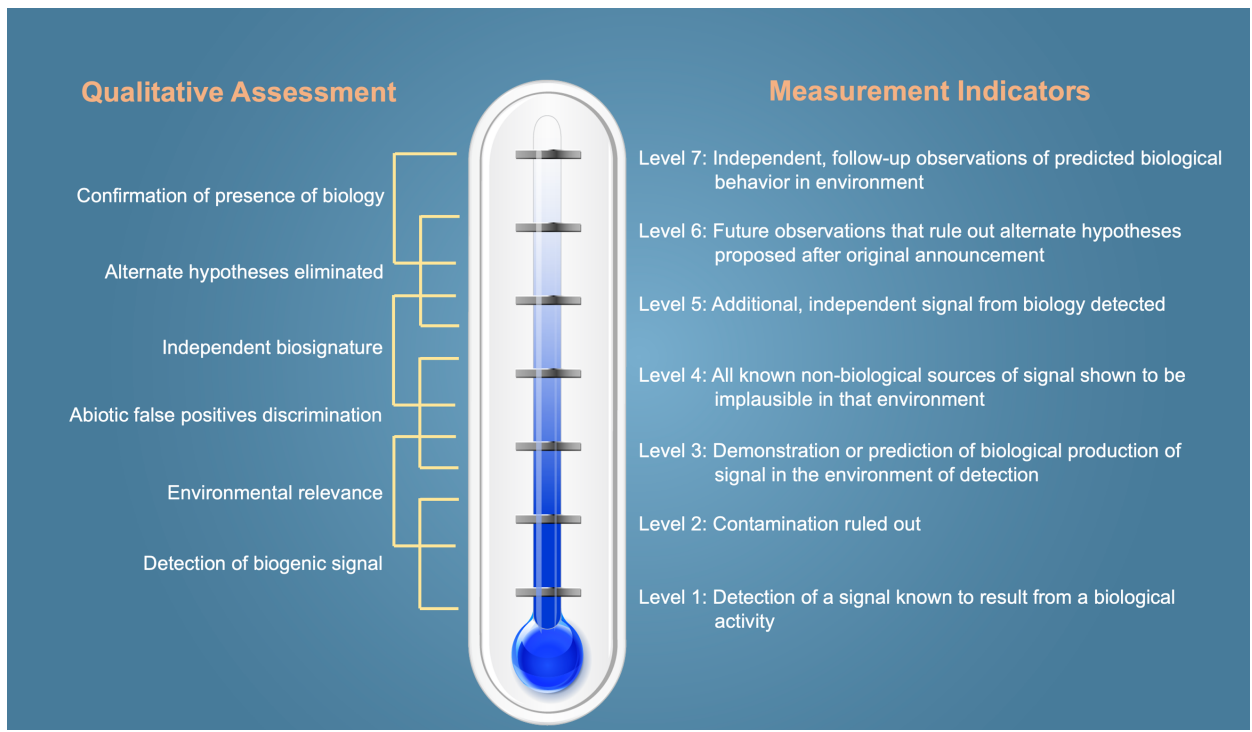

Figure 1. An example Confidence of Life Detection (CoLD) scale.